# ALMA–CTAO Synergies

Paola Andreani[1]
María Díaz Trigo[1]

[1] ESO

**This article aims to raise awareness within the ESO astronomical community of the scientific opportunities offered by the upcoming Cherenkov Telescope Array Observatory (CTAO), highlighting cases where combining the millimetre/submillimetre capabilities of the Atacama Large Millimeter/submillimeter Array (ALMA) with the CTAO's very-high-energy (~TeV) observations enables new avenues for discovery. Many astrophysical sources and environments exhibit emission processes detectable in both regimes; however, the synergies extend beyond shared emission mechanisms and include broader physical processes and observational connections linking millimetre/submillimetre and TeV energies.**

## Motivation

Construction of the Cherenkov Telescope Array Observatory (CTAO) is progressing at the southern site near the Paranal Observatory, hosted by ESO. Together with the northern site in La Palma, Spain, the CTAO will provide full-sky coverage and unprecedented sensitivity for astronomical observations at very high energies (20 GeV – 300 TeV; with 1 TeV = $1.25 \times 10^{-9}$ nm = $2.4 \times 10^{17}$ GHz), improving by at least an order of magnitude over current instruments (Hoffmann & Zanin, 2023). With the commencement of science verification, early science and subsequent full operations, ESO Member States will have access to 10% + 10% guaranteed observing time at the northern and southern CTA sites, respectively.

This suggests that the ESO community should begin preparing projects that make use of the CTAO, with ESO encouraging — though not exclusively — the submission of proposals that exploit synergies with other ESO facilities. Building such a community requires increasing the awareness of high-energy astrophysics, much as ESO did successfully for the Atacama Large Millimeter/submillimeter Array (ALMA), whose sensitivity revolutionised millimetre astronomy; the sensitivity leap offered by the CTAO will play a similar role at high energies.

Given the shift toward a genuinely multi-wavelength and multi-messenger research landscape (for example, Mészáros et al., 2019; Murase & Bartos, 2019; Padovani, 2024), leveraging synergies with ESO facilities will be essential for realising the full scientific potential of the CTAO. While the time-domain community has long operated in this coordinated framework, it has also faced challenges — especially for time-critical observations — as a result of limited community awareness of the operational requirements (for example, Middleton et al., 2017; Díaz Trigo, Maccarone & Tetarenko, 2024). This has motivated the development of Joint Programmes across facilities, with high-energy observatories such as Chandra and XMM-Newton (Schartel & Santos-Lleo, 2025) adopting such approaches early, having recognised the potential for additional scientific output through the use of complementary facilities, especially for rapid multi-wavelength follow-up of energetic, variable events. Strengthening community preparedness and awareness will therefore be fundamental for enabling timely responses in the era of time-domain astronomy.

This article highlights selected scientific cases in which the combined use of the CTAO and ALMA can significantly advance our understanding across many areas of astronomy, potentially leading to major breakthroughs in physics and astrophysics. It also outlines the operational steps needed to fully exploit these synergies.

## Physical processes in play

Gamma rays ($\gamma$-rays) are produced by several physical processes which are categorised in two groups: leptonic processes — thermal bremsstrahlung, inverse Compton scattering, synchrotron emission — and hadronic processes, resulting from inelastic collisions with interstellar material, the dominant emissions being the decay into two $\gamma$-rays of the $\pi^0$ meson ($\pi^0 \rightarrow \gamma\gamma$) and the inelastic proton-proton interactions via the decay chain of charged $\pi$ ($\pi \rightarrow \mu + \nu_\mu \rightarrow e\ \nu_e$).

Each of the different processes has certain characteristics that can be used to identify the underlying production mechanism once the $\gamma$-rays are observed. It is relevant to notice here that the concomitant detection of neutrinos would indelibly mark the hadronic nature of the emission process.

In this article we focus on astrophysical sources and environments where non-thermal emission mechanisms such as synchrotron radiation are detectable at both high ($2 \times 10^{10} - 3 \times\times 10^{14}$ eV) and low ($10^{-4} - 3 \times 10^{-3}$ eV) energies, i.e. targets of the CTAO and ALMA, or where the high-energy radiation influences processes at lower energies via ionisation. In this article we will focus our attention on a few examples that are of particular relevance (in the authors' view).

## Science opportunities

### The origin of cosmic rays, their effects on the initial conditions of star formation, the initial mass function

Cosmic rays (CRs) are a population of non-thermal, relativistic charged particles that pervade the interstellar medium (ISM) of the Milky Way and external galaxies. They are produced primarily in supernova remnants and accelerated by shocks.

The collapse of gas and the onset of star formation within dense, dark molecular clouds — regions where ultraviolet and optical photons cannot penetrate — are regulated by CRs. In these deeply dust-enshrouded environments, CRs serve as the dominant ionisation source, driving the chemistry, setting the gas temperature and facilitating coupling with magnetic fields. Consequently, a close connection is expected between a galaxy's star formation activity and its CR content. In principle, this relationship can be indirectly constrained through observations of the non-thermal emission processes generated by CRs, which span a broad range of frequencies. The millimetre and radio bands trace the CR electron population via synchrotron radiation. However, reliance on these bands alone is challenging, as radio emission depends sensitively on the properties of the galactic magnetic field. In contrast, combining γ-ray observations with simultaneous measurements of star formation tracers provides a powerful and complementary framework for investigating the star formation process.

One of the most original proposals to investigate this issue is the combined measurement of the γ-ray flux and spectrum of CRs in star-forming regions — specifically in dense molecular clouds — using the CTA, together with observations of emission lines from various isotopologues of carbon monoxide ($^{12}C^{16}O$ or CO), $^{13}CO$, and $C^{18}O$, obtained with ALMA. These isotopologues trace the relative $^{13}C$ and $^{18}O$ abundances produced by successive generations of stars (i.e., Zhang et al., 2018a, b).

In very dense molecular clouds, as well as in the extreme environments of compact starbursts within merging galaxies, the CR energy density is expected to be up to a thousand times higher than that in the Milky Way. Under such conditions, CRs can substantially alter the fragmentation of molecular hydrogen ($H_2$) clouds, leading to the formation of fewer low-mass (< 8 $M_\odot$) stars and resulting in a top-heavy stellar initial mass function (IMF). A systematic variation in isotopologue abundance ratios across galaxy-scale molecular hydrogen reservoirs would indicate a departure from the canonical IMF observed in normal star-forming galaxies, suggesting that CR-regulated initial conditions for star formation naturally influence the shape of the stellar IMF (Papadopoulos, Thi & Viti, 2004; Papadopoulos & Thi, 2013).

Related to this topic is the influence of CRs on the evolution of protoplanetary discs. CRs have long been predicted to be the dominant source of ionisation at the midplane of the inner disc (for example, Gammie, 1996). However, ALMA studies of molecular spatial distributions in disc samples suggest that the ionisation rate may be lower than that of the ISM and may vary among discs of similar mass ranges, possibly owing to differences in magnetic field configurations or turbulence (for example, Aikawa et al., 2021). This result could be further explored through direct measurements of CR ionisation rates in discs.

## The Galactic disc and centre

A survey of the Galactic plane is one of the key scientific projects planned by the CTA science collaboration (Abe et al., 2024). The Galactic plane is expected to host a large population of very high-energy sources — pulsar wind nebulae, young and interacting supernova remnants (SNRs), compact binary systems — as well as diffuse emission from CR interactions. In these sources, γ-rays are produced through inelastic interactions between CRs, accelerated by the SNR shock, and the surrounding matter in the molecular cloud (via $\pi^0$ decay). The H.E.S.S. Collaboration (2018) has shown that the diffuse γ-ray emission from the Galactic centre (GC) is spatially coincident with the dense, molecular clouds traced by the CS molecule, as detected by the Nobeyama 45-metre dish (Tsuboi, Handa & Ukita, 1999) and later by tens of observations with ALMA also targeting other gas tracers (Miyawaki et al., 2021 and references therein), extending over a projected distance of 140 pc before fading beyond that point (H.E.S.S. Collaboration, 2018). This γ-ray emission provides strong evidence for a CR source located in or near the GC, with the CR energy profile consistent with a continuous accelerator operating within the central 200 pc.

Complementary ALMA programmes (for example, Sano et al., 2020) have targeted supernova remnants and found evidence for γ-ray production at shock fronts along molecular-cloud edges, consistent with shock–CR interactions. These results may indicate that CRs contribute significantly to the local energy density, influencing the chemistry and physics of star-forming regions.

The additional information provided by ALMA polarisation measurements will help to constrain the magnetic field structure and hence the diffusion of CRs. By mapping how dust emission is polarised, ALMA gives a direct look at field orientation and relative strength. This information is key to inferring magnetic field geometry to the level of scattering and turbulence that shape cosmic ray diffusion. This tightens the physical model and reduces uncertainty in how particles spread from their sources.

## Relativistic jets and shocks associated with accretion around neutron stars, black holes and explosive events

Relativistic jets are highly collimated streams of plasma ejected at relativistic speeds from the vicinity of very compact astrophysical objects such as accreting neutron stars and stellar-mass black holes in X-ray binaries or supermassive black holes. Despite extensive studies performed in the past decades, their launching and quenching mechanisms remain unknown. Jets are expected to be sites of extreme particle acceleration and therefore TeV emitters. However, although most TeV sources are active galactic nuclei (AGN), with flux-limited samples dominated by blazars, AGN in which we are looking directly into the relativistic jet[1], the exact sites of particle acceleration remain unknown. Combined millimetre and TeV observations promise to shed light on this topic. In AGN, millimetre VLBI observations allow the region closest to the black hole to be mapped, including the jet launching area (The Event Horizon Telescope Collaboration, 2019; Lu et al., 2023). Simultaneous observations at TeV energies can then constrain the site of TeV emission by using temporal correlations within millimetre and TeV variability (Akiyama et al., 2015; Algaba et al., 2024) and may extend the relation observed between millimetre and GeV γ-rays for some sources (for example, Leon-Tavares et al., 2011; Ramakrishnan et al., 2016; Kim et al., 2024) to even higher energies. Interestingly, in X-ray binaries, the TeV emission sites have recently been localised to structures in the lobes, far from the centre of the system where the jets are formed (Abeysekara, 2018; LHAASO Collaboration, 2024), and associated at least in two cases with jet termination shocks associated with jet–ISM interaction observed in radio or millimetre emission (Gallo et al., 2005; Tetarenko et al., 2018).

Relativistic ejecta are also observed in long-duration γ-ray bursts (GRBs), bright flashes of γ-rays associated with stellar core collapse events and followed by fading afterglow emission caused by the interaction of the relativistic ejecta with surrounding gas. In the basic picture (see Figure 1), this interaction produces a forward shock in which electrons can be accelerated to relativistic speeds and then emit synchrotron radiation. Temporal evolution of the spectral energy distribution sets important constraints on this model and, in particular, radio and millimetre data have revealed the need to include an additional synchrotron component that is due to a reverse shock propagating back into the jet (Laskar et al., 2013, 2019). Emission at TeV energies has so far been detected only from a handful of GRBs, with the most energetic photon reaching 13 TeV (in GRB 221009A; LHAASO Collaboration, 2023; Abe et al., 2025a). However, these few events already show that TeV emission cannot be simply explained by inverse Compton emission from particle acceleration at the forward shock, suggesting that emission from the reverse shock may be needed (Laskar et al., 2023; Zhang et al., 2024). These models also indicate that the observed evolution of the reverse shock signature at radio/millimetre wavelengths is sensitive to the angular structure of the jet's energy and speed, with further implications for the expected TeV emission from such outflows (Zhang et al., 2024, 2025). Simultaneous monitoring from radio to γ-rays is required to disentangle the two potential emission components, and early radio/millimetre follow-up (within hours to days) is fundamental for capturing and characterising the reverse shock emission.

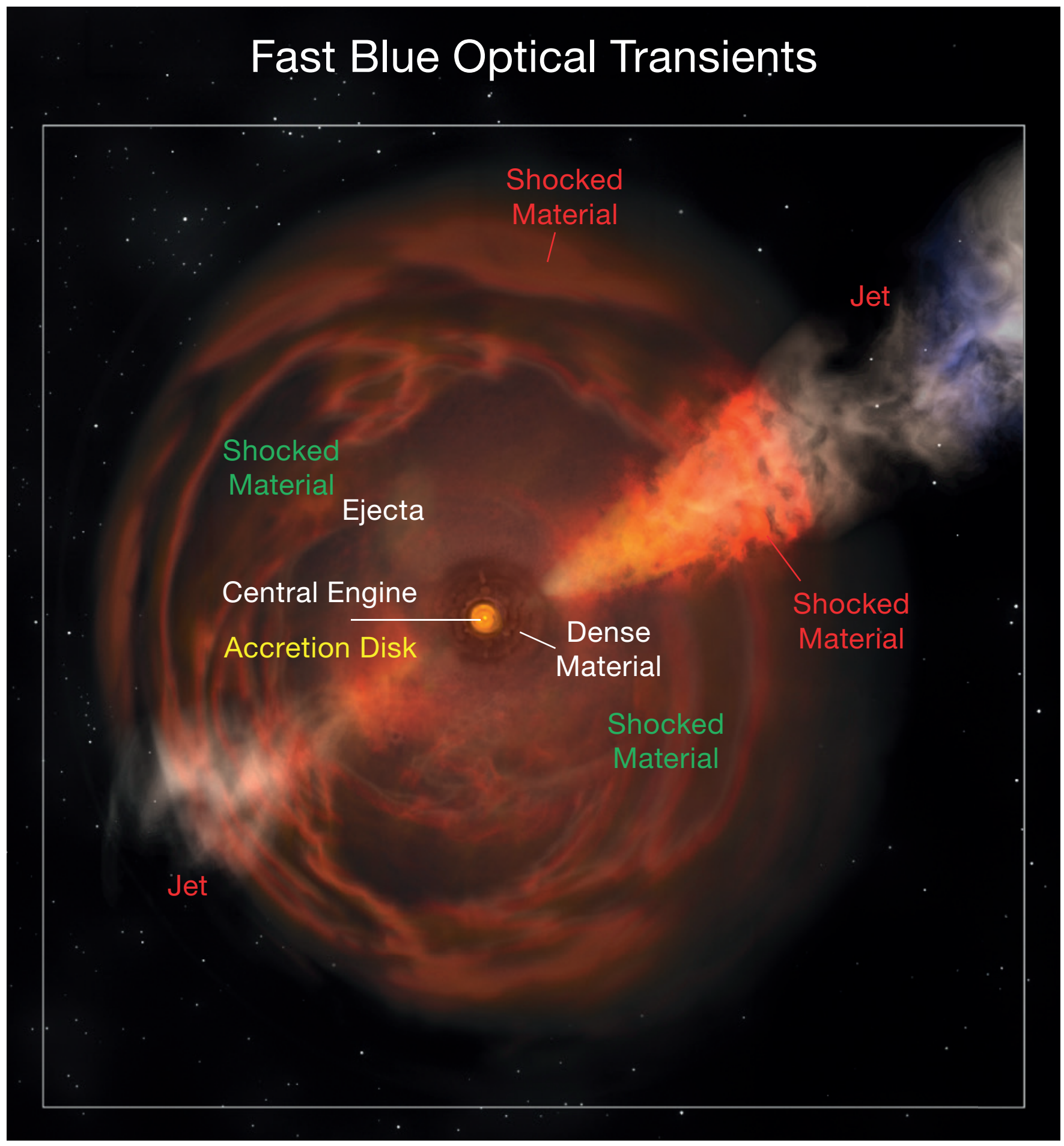

Figure 1. An artist's impression of relativistic ejecta associated with stellar core collapse events and followed by fading afterglow emission, caused by the interaction of such relativistic ejecta with surrounding gas. This simplified picture (adapted from an NRAO public image[2]) shows a forward shock accelerating relativistic electrons, which in turn emit synchrotron radiation. In objects like GRBs, radio and millimetre data suggest an additional synchrotron component that is due to a reverse shock propagating back into the jet . TeV emission also seems to suggest the presence of a reverse shock in addition to the inverse Compton emission from particle acceleration at the forward shock (see text for details).

Besides being sites of particle acceleration in general, shocks are also known to be sites of dust production, best observed via infrared (IR) and millimetre observations. A link between both may be being observed in novae or thermonuclear eruptions on the surfaces of white dwarves in binary systems. Gamma-rays up to TeV energies have recently been detected in the recurrent nova RS Ophiuchi (Acciari et al., 2022; H.E.S.S. Collaboration, 2022; Abe et al., 2025b), representing the first detection of TeV emission in this class of sources. While dust formation in novae has long been established through IR emission accompanied by optical dips that are due to obscuration, the sites of formation are less clear. However, correlations found in a subset of novae with double-peak radio lightcurves between the first peak attributed to synchrotron emission and an optical dip indicating dust formation could now indicate that shocks are the common site of dust production (Derdzinski, Metzger & Lazzati, 2017) and particle acceleration (Chomiuk et al., 2021). Simultaneous IR/millimetre and γ-ray monitoring observations, together with high-spatial-resolution IR/millimetre observations could help confirm this hypothesis.

Finally, millimetre observations are now also helping to elucidate the origin of γ-rays in the so-called γ-ray binaries, a sub-class of binaries with a massive star orbiting a compact object and broadband non-thermal emission peaking above 1 MeV. As an example, multi-band ALMA observations of the nearby source PSR B1259-63/LS 2883, consisting of a pulsar in an eccentric orbit around a Oe-type companion star with an equatorial decretion disc, have been crucial in disentangling the changes in synchrotron emission at low frequencies and the circumstellar disc at high frequencies as the pulsar passes through or near the disc of the companion star, when the interaction of the pulsar wind and the stellar environment leads to enhanced high-energy emission (Fujita et al., 2024).

The significant increase of sensitivity offered by the CTAO should also enable the detection of TeV energies from additional transient and variable sources (Abe et al., 2025c), thus further complementing studies of the non-thermal emission at radio and millimetre wavelengths, which will remain key for locating the acceleration sites and shocks.

## Challenges

Operational models built for traditional astronomy struggle because they do not account for the fast reaction and tight coordination required today; communication often lags and events can fade away before observatories can respond. Technical systems cannot always keep up with the physical timescales involved. Data reduction adds more friction, especially when teams must work through large, uneven datasets.

Ground-based telescopes face visibility limits that vary by site, which makes simultaneous coverage difficult; space assets can run into the same problem. Time allocation procedures were never designed for rapid multi-messenger campaigns, so observing time rarely lines up with the needs of the science. Data processing still leans heavily on specialists in each wavelength, and archives do not always provide the key physical quantities in forms that are easy to use or cross compare.

Strong coordination with multi-messenger facilities is essential. That includes neutrino detectors like IceCube and gravitational wave observatories such as the Laser Interferometer Gravitational wave Observatory (LIGO), Virgo, the Kamioka Gravitational Wave Detector (KAGRA), the Laser Interferometer Space Antenna (LISA) and the planned Einstein Telescope.

The first steps toward modernising operational models are straightforward. Large facilities need to join the fast alert networks already used across the multi-messenger community. They also need joint observing programmes with Target of Opportunity capabilities that allow quick action when an event appears. Facilities must coordinate schedules in real time so they can follow fading sources at the same moment, not hours later. Alongside this, data reduction tools must be built for highly variable objects and should provide light curves and other essential products in real time. These changes create the foundation for facilities to work together and fully exploit the scientific power of synergetic observations.

## Outlook

The synergetic use of astronomical facilities enables tackling some of the scientific questions that are unreachable by a single facility, thereby opening the parameter space of scientific discovery. A key factor for a successful synergetic use of two (or more) facilities is the matching of some technical capabilities such as sensitivity or angular resolution or their complementarity, for example spectroscopy and imaging. The second key factor is the complementarity of the facilities for the study of physical processes. In this article we have given an indication of the scientific cases that we expect to open new avenues for discovery when investigated with ALMA and the CTA. We also point to some operational changes that are needed for realising this potential, particularly — although not exclusively — in the cases that involve highly variable phenomena, and encourage ALMA and the CTAO to start a discussion now on how to implement such changes to be ready when the CTA comes online.

### Acknowledgements

M. Díaz Trigo thanks T. Hovatta and T. Laskar for helpful discussions on the ALMA–CTA synergies in respect of AGN and GRBs. P. Andreani thanks P. Papadopulos for having raised the issue of the CO isotopologues during one of the scientific discussions at ESO.

### Links

[1] TeV catalogue: https://www.tevcat.org/
[2] NRAO public image: https://public.nrao.edu/gallery/nrao20df03b/